\documentclass{llncs}
\usepackage[T1]{fontenc}
\usepackage{amsmath}
\usepackage{graphicx}
\usepackage{lineno,hyperref}
\usepackage{colortbl}
\usepackage{amsmath,amssymb,amsfonts}
\usepackage{algorithmic}
\usepackage{graphicx}
\usepackage{float}

\begin{document}

\title{Reconstruct Spine CT from Biplanar X-Rays via Diffusion Learning}

\titlerunning{Diff2CT}


\author{
Zhi Qiao\inst{1} \and 
Xuhui Liu\inst{1, 3} \and
Xiaopeng Wang\inst{1, 4} \and 
Runkun Liu\inst{1} \and
Xiantong Zhen\inst{1} \and 
Pei Dong\inst{2} \and
Zhen Qian\inst{1}
}

\institute{Beijing United Imaging Research Institute of Intelligent Imaging, Beijing, China
\and United Imaging Intelligence, Beijing, China
\and Beihang University, Beijing, China 
\and Tianjin University, Tianjin, China}

%
%
\maketitle              
\begin{abstract}
Intraoperative CT imaging serves as a crucial resource for surgical guidance; however, it may not always be readily accessible or practical to implement. In scenarios where CT imaging is not an option, reconstructing CT scans from X-rays can offer a viable alternative. In this paper, we introduce an innovative method for 3D CT reconstruction utilizing biplanar X-rays. Distinct from previous research that relies on conventional image generation techniques, our approach leverages a conditional diffusion process to tackle the task of reconstruction. More precisely, we employ a diffusion-based probabilistic model trained to produce 3D CT images based on orthogonal biplanar X-rays. To improve the structural integrity of the reconstructed images, we incorporate a novel projection loss function. Experimental results validate that our proposed method surpasses existing state-of-the-art benchmarks in both visual image quality and multiple evaluative metrics. Specifically, our technique achieves a higher Structural Similarity Index (SSIM) of 0.83, a relative increase of 10\%, and a lower Fréchet Inception Distance (FID) of 83.43, which represents a relative decrease of 25\%.

\end{abstract}

\section{Introduction}


Intraoperative X-ray is a commonly used projectional imaging technique during spine surgeries, highly valued for its capacity to deliver outstanding bone contrast and facilitate quick assessment of the placement and condition of surgical implants like screws. When compared to other imaging technologies, X-ray presents multiple advantages such as reduced radiation dosage, cost-efficiency, and quicker imaging times. Nevertheless, it has a significant limitation: it cannot furnish a complete three-dimensional view of internal anatomical structures. In contrast, 3D imaging modalities, including computed tomography (CT), provide a more nuanced and comprehensive perspective of the targeted area. This in-depth visualization aids surgeons in accurately identifying the location and morphology of various structures, like bones and implants, thereby contributing to more effective intraoperative assessment. Hence, when CT imaging is either impracticable or inaccessible, reconstructing CT scans from X-ray images stands as a potentially invaluable substitute.


Several previous studies have made attempts at reconstructing 3D images from 2D X-ray images. In particular, some researchers proposed using Convolutional Neural Networks (CNNs) in \cite{1Singleimage,3knee1xray} to predict a volume representation of the underlying 3D object from a single X-ray projection view. However, due to the limited information contained in a single 2D X-ray image, such approaches may not be able to achieve high accuracy in 3D reconstruction. Several recent studies have shown that using biplanar X-ray inputs can be advantageous for CT reconstruction, as demonstrated in \cite{5X2CT,4knee2xray,6CTRSNet}. For instance, researchers in \cite{4knee2xray} proposed an end-to-end deep network for the 3D reconstruction of knee bones, while in \cite{6CTRSNet}, a multi-scale fusion and hierarchical reconstruction framework was proposed for head and neck CT reconstruction. Among these studies, GAN-based reconstruction methods were found to achieve the best visual image quality \cite{5X2CT}.

The Denoising Diffusion Probabilistic Model (DDPM) \cite{diffusion_ddpm} has demonstrated remarkable performance in generating realistic images when compared to other generative models such as Flow-based models, Variational Autoencoders (VAE)\cite{vae}, and Generative Adversarial Networks (GANs)\cite{7conditiongan}. The model achieves this by progressively denoising a normally distributed variable, which entails acquiring knowledge of the inverse process of a Markov Chain of fixed length $T$, thereby learning the distribution $p(x)$ of the data. Since the DDPM is a likelihood-based model, it does not suffer from mode-collapse or training instability, as is often the case with GANs.


In this paper, we leverage the benefits of diffusion models in image generation to address a clinically significant challenge in intraoperative imaging: CT reconstruction from X-rays. We frame CT reconstruction from X-rays as a conditional generative learning problem, following prior work in the field \cite{diffusion_condition_ddpm,diffusion_highreso_ddpm}. In this framework, intraoperative X-ray images serve as condition information. Thus, we introduce a novel diffusion model-based CT reconstruction approach, which we call \textbf{Diff2CT}. Our method utilizes a conditional denoising autoencoder to predict a denoised version of the CT image. This process is controlled through the input orthogonal biplanar X-ray images. Additionally, we introduce a novel projection loss term to enforce structural consistency in 3D space. Finally, we evaluate the performance of Diff2CT on a real-world lumbar pedicle screw fixation dataset. The main contributions of this work can be summarized as follows:

\begin{itemize}
   \item Our study represents the pioneering attempt to explore CT reconstruction utilizing diffusion models from bi-planar X-rays. The proposed approach leverages the structural information of underlying organs in bi-planar X-rays and models conditional distributions for denoised variant regression. 
   \item Compared to extant reconstruction algorithms, our approach exhibits superior quantitative performance and yields the highest quality of visual image output.
    
\end{itemize}

\section{Proposed Method}

Diffusion models are probabilistic models designed to learn the distribution $p(x)$ of data by gradually denoising a normally distributed variable. This process corresponds to learning the reverse process of a fixed Markov Chain of length $T$. In the context of image synthesis, diffusion models utilize a reweighted variant of the variational lower bound on $p(x)$, which resembles denoising score-matching. These models can be interpreted as a sequence of equally weighted denoising autoencoders $\epsilon_{\theta}(x_t, t )~t=1...T$, which are trained to predict a denoised variant of their input $x_t$. The input $x_t$ corresponds to a noisy version of the original input $x$. The corresponding objective function can be expressed as follows,
\begin{equation}
    L_{DM}=E_{x, \epsilon \sim \mathcal{N}(0, 1),t}\left [ \left \| \epsilon - \epsilon_{\theta }(x_t, t) \right \|^2_2  \right ] 
\end{equation}



The original DDPM is an unsupervised learning method that is inadequate for generating images with desired semantics since it is intended for unconditional image generation. However, CT reconstruction from biplanar X-rays is a conditional generation task where biplanar X-ray inputs can provide a more comprehensive and detailed representation of the anatomy and structures of interest. In the case of implants, bones, and other organs, the generated CT images must be consistent with the position and morphology of the biplanar X-rays, rather than a random sample in the target CT distribution. Inspired by certain conditional diffusion methods, our conditional DDPM decoder takes in ($x_t, \mathcal{X}_1, \mathcal{X}_2$) as inputs to produce the output image, where $\mathcal{X}_1$ and $\mathcal{X}_2$ are the biplanar X-ray inputs. The decoder models $p_\theta(x_{t-1}|x_t, \mathcal{X}_1, \mathcal{X}_2)$ to match the inference distribution $p_\theta(x_{t-1}|x_t, x_0)$. By introducing the condition $\mathcal{X}_1$ and $\mathcal{X}_2$ in the reverse process, the noise estimator $\epsilon_{\theta }(x_t, t)$ in DDPM becomes $\epsilon_{\theta }(x_t, t, \mathcal{X}_1, \mathcal{X}_2 )$. Finally, the noise-prediction loss function can be modified as follows,




\begin{equation}
\label{eq:reloss}
    L_{DM}=E_{x, \epsilon \sim \mathcal{N}(0, 1),t}\left [ \left \| \epsilon - \epsilon_{\theta }(x_t, t, \mathcal{X}_1, \mathcal{X}_2  ) \right \|^2_2  \right ] 
\end{equation}
\\
where $\epsilon_{\theta }(x_t, t, \mathcal{X}_1, \mathcal{X}_2  )$ is the output of X2NoiseNet.

\subsection{X2NoiseNet}

In contrast to existing conditional noise prediction networks, our approach involves mapping 2D data to 3D space while preserving most of the spatial information that is critical for restoring regions of interest in 3D CT data. To accomplish this, we draw inspiration from view alignment based dual X-ray image merging techniques \cite{4knee2xray,5X2CT} and employ a matrix permutation process to maintain geometric consistency in orientation for the biplanar features, followed by a direct addition operator for further fusion. Specifically, our model requires that the condition inputs ($ \mathcal{X}_1$ and $ \mathcal{X}_2$) be first expanded directly from the 2D data to 3D by repeating the 2D data along the respective X-ray projection directions in a three-dimensional space.  This procedure is based on the assumption commonly used by back-projection-based CT reconstruction methods that a 2D X-ray image represents an integration of the 3D volume along the direction of projection, and therefore the density of the volume can be simplified to be uniformly distributed along the projection lines, particularly when there is an absence of prior knowledge. In order to transform these inputs into a unified coordinate space, we align the two 3D features in a consistent orientation, as follows.

\begin{equation}
\begin{split}
&Feat3D(\mathcal{X}_1): [y, z, x, ...] \overset{Permute}{\rightarrow} [x, y, z, ...]  \\
&Feat3D(\mathcal{X}_2): [x, z, y, ...] \overset{Permute}{\rightarrow} [x, y, z, ...] 
\end{split}
\end{equation}
where $Feat3D(\cdot)$ represents the 3D feature, with $x$, $y$, $z$ representing different axes, and $\dots$ representing the left channel features. To combine the two 3D features as condition features, the addition operator is utilized, yielding $Feat3D( \mathcal{X}_1) + Feat3D( \mathcal{X}_2)$. Subsequently, the noisy image $x_t$ and condition features are combined using the concatenation operator, followed by a time-conditional UNet, as described in \cite{diffusion_ddpm}. The detailed noise prediction network is depicted in Figure \ref{method:framework}.

\subsection{Loss Functions}

First, we formulate a reconstruction loss function that serves as a means to ensure that the reconstructed noise is closely aligned with the ground truth at the voxel level, as represented in Eq.\ref{eq:reloss}. Drawing inspiration from \cite{3dprojection}, we leverage 2D projections to the predicted volume and compare the resulting projection views with the projection images from the corresponding ground truth. To streamline the process and focus more on general shape consistency, rather than accuracy, we employ orthogonal projections instead of perspective projections. Specifically, we employ three orthogonal projection planes, namely the axial, coronal, and sagittal planes. We define the reconstruction loss as follows.


\begin{equation}
\begin{split}
L_{re}(G) = & E_{x, \epsilon \sim \mathcal{N}(0, 1),t} [ ||\epsilon_0-\epsilon_\theta (x_t, \mathcal{X}_1, \mathcal{X}_2  )||^2 + \\
& \frac{1}{3}(||\mathcal{P}_A(\epsilon_0)-\mathcal{P}_A(\epsilon_\theta (x_t, \mathcal{X}_1, \mathcal{X}_2  ))||^2 \\
& + ||\mathcal{P}_C(\epsilon_0)-\mathcal{P}_C(\epsilon_\theta (x_t, \mathcal{X}_1, \mathcal{X}_2  ))||^2 \\
& + ||\mathcal{P}_S(\epsilon_0)-\mathcal{P}_S(\epsilon_\theta (x_t, \mathcal{X}_1, \mathcal{X}_2  ))||^2)]
\end{split}
\end{equation}

where $\mathcal{P}_A(\cdot)$, $\mathcal{P}_C(\cdot)$, $\mathcal{P}_S(\cdot)$ denote the axial, coronal, and sagittal projections of the object.

\section{Experiments}

In this section, we introduce a real-world dataset of CT images post lumbar pedicle screw insertion operations, followed by an evaluation of the proposed Diff2CT model using several established metrics. To demonstrate the efficacy of our method, we compare it with state-of-the-art approaches, providing fair comparisons and a comprehensive analysis to showcase the improvements achieved by our proposed model over the baselines. Lastly, we present the CT reconstruction results from a set of real-world intraoperative biplanar X-rays, demonstrating the potential and practicality of Diff2CT.

\begin{figure*}[h]
\centering
\includegraphics[width=0.85\textwidth]{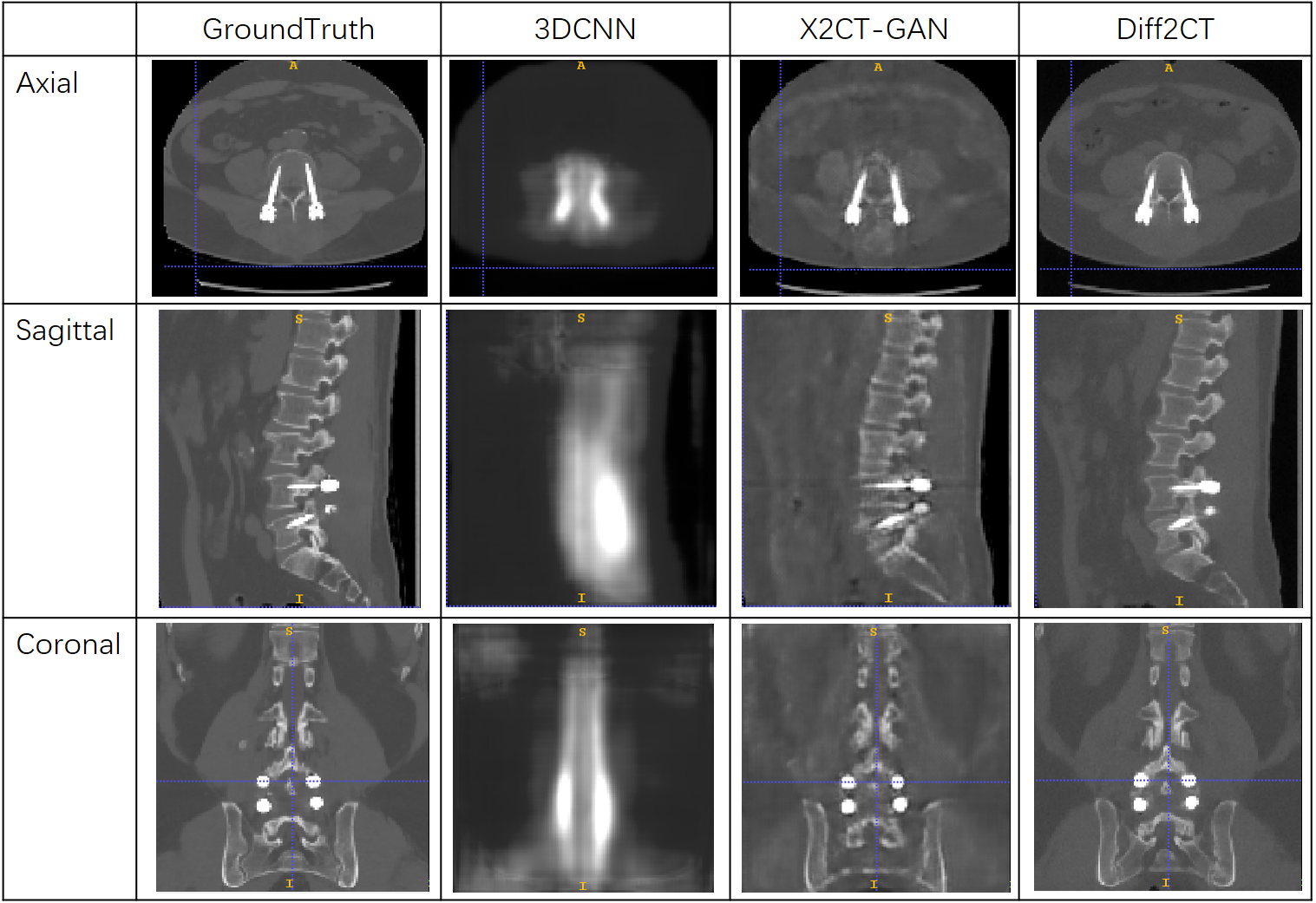}
\caption{Representative CT reconstruction results shown in the mid-axial (1st row), mid-sagittal (2nd row), and mid-coronal (3rd row) views. Our method is compared with baseline methods (PSR, 3DCNN, Diff2CT) and the ground truth (GT).}
\label{exp:qualitative}
\end{figure*}

\subsection{Datasets}

In our experiments, we constructed a lumbar vertebra dataset post pedicle screw insertion operations comprising 268 3D CT scans. To establish a training set, we randomly selected 212 CT scans while reserving the remaining 56 CT scans for testing. To ensure consistency, we initially resampled all CT scans to an isotropic voxel resolution of $2\times 2 \times 2 ~ mm^3$ and then center-cropped them to a fixed size of $128 \times 128 \times 128$.

Following this, we employed CT value clipping to limit the CT value range to [-1000, 4096], and min-max normalization to scale data to the [-1, 1] range. While ideally, we would have preferred a paired dataset of both CT and biplanar X-ray data for training and testing our proposed model, such data proved challenging to obtain, and we had to resort to synthesizing the corresponding X-rays from volumetric CT images using digitally reconstructed radiographs (DRR) technology \cite{13ddr} to generate the required paired CT and X-ray images. Subsequently, we used CycleGAN \cite{7conditiongan} to learn the mapping from synthetic to real X-rays, transforming all DRR-based synthetic X-rays into more realistic X-rays, as previously demonstrated in \cite{5X2CT}.


\begin{table*}
\centering
\caption{Quantitative results of baseline methods and Diff2CT(The values in parentheses are the standard deviations.)}
\begin{tabular}{|l|c|c|c|c|}
\hline 
Model         &             MAE &          PSNR &  SSIM $\uparrow $&         FID $\downarrow $ \\ 
\hline
PSR           & 0.03258(8.9e-5) & 25.1421(2.246)& 0.6025(0.002)& 256.1991(29.66)   \\
\hline 
3DCNN         & 0.02982(9.1e-5) & 25.4031(2.208)& 0.6328(0.001)& 237.4457(24.06)   \\
\hline 
X2CT-GAN      & \textbf{0.01975(4.6e-5)} & 27.8361(2.947)& 0.7673(0.006)& 123.6737(36.51) \\
\hline
Diff2CT       & 0.05916(6e-4) & \textbf{27.8391(6.542)}& \textbf{0.8318(0.003)}& \textbf{83.4372(29.32)} \\
\hline
\end{tabular}
\label{table:measurement}
\end{table*}

\begin{figure*}
\begin{center}
\begin{minipage}{0.47\textwidth}
\includegraphics[width=2.3in, height=1.5in]
{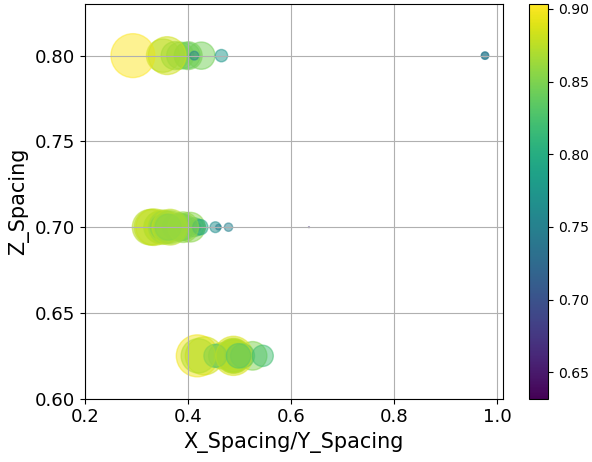}
\caption{Analysis of results (A lighter color indicates a larger SSIM value, and a larger diameter of the circle indicates a larger PSNR value)}
\label{exp:distribution}
\end{minipage}
\begin{minipage}{0.47\textwidth}
\includegraphics[width=2.3in, height=1.5in]
{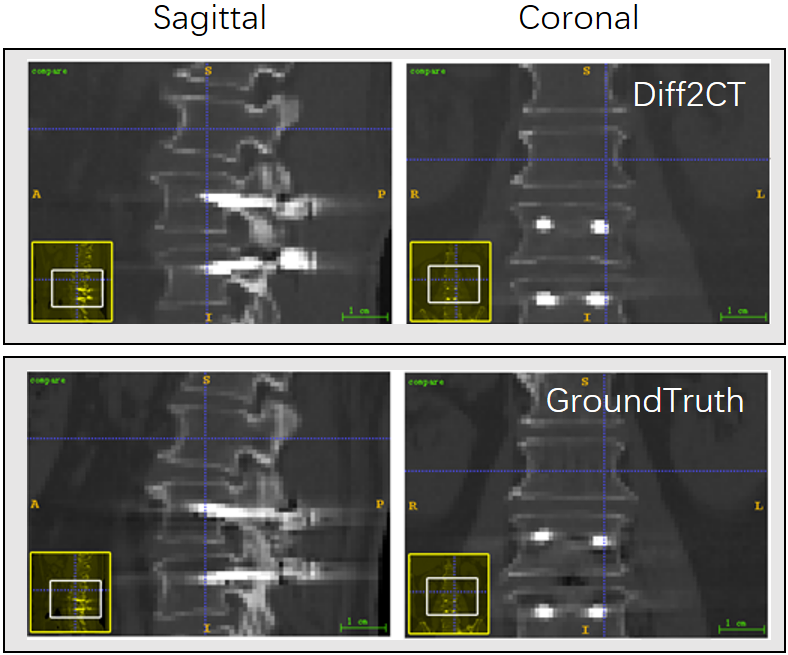}
\caption{A case for low resolution CT image with original pixel spacing of  [0.9765,0.9765,0.8].  In preprocessing, we firstly resample all of CT images into a unified [2,2,2] pixel spacing.}
\label{exp:spacing_discuss}
\end{minipage}
\end{center}
\end{figure*}

\subsection{Settings}

To implement the denoising module, we utilized the network architecture designed in DDPM \cite{diffusion_ddpm}. The noise level was varied from $10^{-6}$ to $10^{-2}$ in a linear manner with a schedule of $t = 1000$. We adopted the Adam algorithm \cite{9adam} with a learning rate of $2 \times 10^{-4}$ and trained the model for 1000 epochs with a batch size of 1. For the CT reconstruction task, we selected three state-of-the-art methods to serve as baselines:


\begin{itemize}
\item[--] PSR \cite{1Singleimage} utilized deep learning-based convolutional neural networks to reconstruct 3D volumes based on a single 2D X-Ray.
\item[--] 3DCNN \cite{4knee2xray} applied encoder-decoder framework to model the mapping from biplanar 2D X-rays to CT.
\item[--] X2CT-GAN \cite{5X2CT} used GAN framework and proposed specific generator to fuse view features in the decoder, where biplanar 2D X-rays were the inputs. 
\end{itemize}

In order to establish a baseline benchmark, we replicated the PSR and 3DCNN models, and utilized the publicly available code of X2CT-GAN. The model parameters were instantiated in accordance with their corresponding publications. 

Various metrics are used to evaluate the quality of synthetic images compared to reference CT images, including the Mean Absolute Error (\textbf{MAE}), the Peak Signal-to-Noise Ratio (\textbf{PSNR}), the Structural Similarity (\textbf{SSIM}) and the Frechet Inception Distance (\textbf{FID}). FID score is a metric that determines the distance between computer vision features of real and generated images using the inception v3 model. FID has the advantage of better matching human subjective evaluations, and it can simultaneously evaluate the realism of the reconstructions and their consistency with observations, setting it apart from other metrics.



\subsection{Qualitative Results}
The CT reconstruction results are first evaluated qualitatively, and the findings are presented in Fig.\ref{exp:qualitative}. It is observed that the use of PSR with only a single-view X-ray leads to blurrier volumes when compared to the other biplanar X-ray-based models. Furthermore, when compared to the 3DCNN, the X2CT-GAN generates sharper boundaries of the organs and implants and exhibits better visual reconstruction quality, highlighting the advantages of GANs in generative tasks. Ultimately, the proposed Diff2CT approach produces the best visual results, with the closest morphological characteristics of the vertebrae and implants to the ground truth and the sharpest boundaries. These results suggest that the proposed method achieves less reconstruction distortion and less detailed texture loss.


\subsection{Quantitative Results}
Detailed quantitative results are presented in Table \ref{table:measurement}. The results indicate that: firstly, the PSR model, when trained with a single X-ray input, exhibits the poorest performance when compared to the other biplanar models. Secondly, the X2CT-GAN model demonstrates superior performance as compared to the 3DCNN model. This suggests that Generative Adversarial Networks (GANs) have the potential to better recover the underlying 3D anatomy. 
Thirdly, the Diff2CT model exhibits higher SSIM with a relative increase of 10\% and lower FID with a relative decrease of 25\% as compared to the best baseline. However, it is observed that the model does not demonstrate a significant improvement in terms of PSNR and even exhibits worse MAE scores. This can be attributed to the fact that all the baseline models are regression models which can achieve lower MAE scores. Additionally, as highlighted in \cite{diffusion_highreso_ddpm}, these metrics may not accurately align with human perception \cite{diffusion_107} and tend to favor blurriness over imperfectly aligned high-frequency details \cite{diffusion_74}.


Additionally, we investigate the correlation between measurement scores and the original pixel spacing values of the test dataset. This relationship is depicted in Figure \ref{exp:distribution}. Our findings reveal that the measurement scores are positively related to the CT resolution. This suggests that our model can produce better reconstruction performance for CT images with smaller pixel spacing values. We postulate that low-resolution images exhibit reduced contrast in high-frequency regions due to resampling during the preprocessing step for pixel space consistency. In these low-contrast regions, our Diff2CT approach can enhance contrast, thereby improving visual image quality. However, this enhancement leads to poorer evaluation results. We present a specific example of this issue in Figure \ref{exp:spacing_discuss}.


\subsection{Case Study}

As the primary objective of this study is to generate a CT scan from real X-rays, we assess the performance of our model on real-world intraoperative X-ray data, despite the fact that the model is trained on synthetic data. Given the lack of corresponding ground truth for these X-ray images, we limit our evaluation to qualitative analysis. Figure \ref{exp:realworld} presents our results, indicating that the reconstructed lumbar vertebra and screws are visually realistic and plausible compared to the real-world biplanar X-rays.




\section{Conclusion}
In this paper, we propose a novel biplanar X-rays-conditioned diffusion probabilistic model for intraoperative spine CT reconstruction.  The experimental results demonstrate that our approach outperforms the SOTA baselines in terms of visual and quantitative CT quality. Specifically, our approach yields 10\% and 25\% improvements in SSIM and FID scores, respectively, when compared to the SOTA baselines.


\bibliographystyle{splncs04}
\bibliography{main}

\end{document}